# Modeling Human Erythrocyte Shape and Size Abnormalities

S. Muñoz San Martín,[1] J.L. Sebastián,[1] M. Sancho[1] and G. Álvarez[2]

[1]*Departamento de Física Aplicada III, Facultad de Ciencias Físicas, Universidad Complutense, 28040 Madrid, Spain*

[2]*Departamento de Física Teórica II, Facultad de Ciencias Físicas, Universidad Complutense, 28040 Madrid, Spain*

We present simple parametric equations in terms of Jacobi elliptic functions that provide a realistic model of the shape of human normal erythrocytes as well as of variations in size (anisocytosis) and shape (poikilocytosis) thereof. We illustrate our results with parameterizations of microcytes, macrocytes and stomatocytes, and show the applicability of these parameterizations to the numerical calculation of the induced transmembrane voltage in microcytes, macrocytes and stomatocytes exposed to an external RF field of 1800 MHz.

**INTRODUCTION**

Erythrocytes are the most numerous cells in the blood and the development of realistic models for their shape has attracted much effort. For example, we have previously shown the crucial role played by the geometry of the cell model in the response of erythrocytes to electromagnetic fields [Sebastian et al, 2004]. The erythrocyte normal shape resembles a biconcave lens. Although erythrocytes are devoid of a nucleus and are terminally differentiated (that is, they never divide), they can undergo a variety of morphological changes associated to various medical disorders [Gedde et al, 1995]. In this paper we address the modeling of abnormal variations in size which maintain the basic biconcave shape of the erythrocyte (anisocytosis), and abnormal variations in shape which maintain the original volume of the erythrocyte (poikilocytosis).

Anisocytosis is the generic term used to describe an abnormally wide distribution of the sizes of the erythrocytes in the blood, be it the presence of young red blood cells such as macrocytes which are larger than mature normal red blood cells, or the presence of smaller red blood cells such as microcytes. Although anisocytosis by itself is not diagnostic, an improved anemia classification might be available by combining measures of red blood cell size variability with mean corpuscular volume [Simel et al, 1988]. Microcytic cells occur in immune-mediated hemolytic anemia, microvascular constriction, early Heinz-body anemia and iron-deficiency anemia. Macrocytic cells occur with regenerative anemia and rarely with erythrocytic leukemia.

The rather non-specific term poikilocytosis is used to describe an unusually high (greater than 10%) population of abnormally shaped erythrocytes in peripheral blood [Harvey, 2001]. These morphologically abnormal erythrocytes (poikilocytes) may be caused by a variety of conditions, including fragmentation of erythrocytes, oxidative injury, immune-mediated damage, and congenital abnormalities [Cowell, 1999]. Among these morphologically abnormal cells are the stomatocytes, characterized by a slit-like zone of central pallor. Stomatocytes are usually the result of a defect in the erythrocyte sodium pump, and particularly interesting because of their abnormal increase in osmotic fragility which could lead to mechanical damage in turbulent blood flow circulation. Poikilocytes are not considered highly sensitive or specific indicators of disease; however, familiarity with common erythrocytic morphological changes may help veterinarians to identify the underlying causes of disease in some animals.

Considering the essential role played by the shape and size of the red blood cells in all the processes mentioned above, in this work we focus on the modeling of the two more extreme cases of anisocytes [Auer, 2005] (macrocytes and microcytes), and of stomatocytes as typical representatives of poikilocytes.

**MODELING ERYTHROCYTE SHAPE ABNORMALITIES**

Among the methods frequently used to model the erythrocyte normal shape [Sebastian et al, 2004, Moon and Spencer, 1998] we mention the surface of revolution generated by a Cassini curve [Gray, 1998]. Its main advantage is that Cassini curves have simple parametric representations in terms of trigonometric functions; its main limitations are that Cassini curves are determined by only two parameters (which can be used to fix only the length and the height of a real erythrocyte), and that the implementation of anisotropic deformations is not straightforward (in particular anisocytosis, poikilocytosis and the changes in the shape of membranes which occur spontaneously when the cells are immersed in an aqueous environment under appropriate conditions [Jie et al, 1998]). To find a reasonably simple parametric representation that allows for the modeling of these deformations we recall the parametric representation of the normal biconcave shell in terms of the $\text{sn}(u,m)$, $\text{cn}(u,m)$ and $\text{dn}(u,m)$ Jacobi elliptic functions with three free parameters given by Kuchel and Fackerell [1999], which we prefer to write directly in terms of as many physically measurable parameters as possible:

$$\mathbf{r}(u,\phi) = \left( \frac{\ell}{2} \text{cn}(u,m) \cos\phi, \frac{\ell}{2} \text{cn}(u,m) \sin\phi, \pm h_0 \, \text{sn}(u,m) \frac{\text{dn}(u,m)}{\text{dn}(U,m)} \right), \qquad (1)$$

where $\ell$ is the diameter of the erythrocyte, $2h_0$ is the height of the erythrocyte at its center, $U = K(m)$ is the corresponding complete elliptic integral of the first kind (which satisfies $\text{cn}(U,m) = 0$, $\text{sn}(U,m) = 1$ and $\text{dn}(U,m) = \sqrt{1-m}$, although for clarity we keep the functional form of this last expression in the parametric equations), $u \in [0, U]$, $\phi \in [0, 2\pi]$, and the plus and minus signs correspond to the upper and lower half of the cell respectively. The only remaining parameter is $m \in [0,1)$, which can be used to fix, for example $2h_{\max}$, the maximum height of the erythrocyte. Equation (1) is obviously suitable to model not only normal erythrocytes but also macrocytes and microcytes.

The main result of this paper is the following: by modifying the third component of Eq. 1 in the form

$$\mathbf{r}_+(u,\phi) = \left( \frac{\ell}{2} \text{cn}(u,m_+) \cos\phi, \frac{\ell}{2} \text{cn}(u,m_+) \sin\phi, h_+ \, \text{sn}(u,m_+) \left( \frac{\text{dn}(u,m_+)}{\text{dn}(U_+,m_+)} \right)^3 \right), \qquad (2a)$$

$$\mathbf{r}_-(u,\phi) = \left( \frac{\ell}{2} \text{cn}(u,m_-) \cos\phi, \frac{\ell}{2} \text{cn}(u,m_-) \sin\phi, -h_- \, \text{sn}(u,m_-) \left( \frac{\text{dn}(u,m_-)}{\text{dn}(U_-,m_-)} \right)^2 \right), \qquad (2b)$$

and by using independent values of the parameters $h_\pm$ and $m_\pm$ (and the ensuing $U_\pm = K(m_\pm)$) for the upper and lower part of the cell we can easily model shape alterations like stomatocytes while keeping all the advantages of a simple, explicit parametric representation.

By way of example, Fig. 1a shows the parameters and 2D generating curves for the basic discocyte and a stomatocyte of the same volume obtained from Eqs. 1 and 2 respectively. Similarly, Fig. 1b shows the 3D cell geometries generated by rotation of these curves around the vertical axis. As it can be easily seen both in the 2D and in the 3D figures, the stomatocyte has almost lost the indentation on the lower part of the cell, and the oval area of central pallor, sometimes referred to as a "mouth", is perfectly modeled by the modified Eq. 2. In Table 1 we present the numerical values of the parameters of these cells as well as of the parameters of the macrocyte and microcyte used in the next section. Note that the maximum height of the stomatocyte has to be determined by using the inverse functions $u_\pm = \mathrm{sn}^{-1}(x, m_\pm)$, and that the volumes of the upper and lower parts of the cells can be readily calculated thanks again to the parametric equations:

$$V_\pm = -2\pi \int_0^{U_\pm} x'(u, m_\pm) x(u, m_\pm) z(u, m_\pm) du, \qquad (3)$$

where the prime denotes derivation with respect to $u$.

**TABLE 1. Parameters used in Eqs. 1 and 2 to model the normal and altered erythrocytes:** $\ell$ **is the diameter of the cell,** $2h$ **(** $h_+ + h_-$ **for the stomatocyte) is the height at the origin,** $m$ **(** $m_+$ **and** $m_-$ **for the stomatocyte) is the parameter in the Jacobi elliptic functions,** $V$ **is the volume,** $2h_{max}$ **the maximum height and** $\delta$ **the thickness of the membrane.**

| Cell | $\ell\,(\mu m)$ | $h\,(\mu m)$ | $m$ | $V\,(\mu m^3)$ | $2h_{max}\,(\mu m)$ | $\delta\,(nm)$ |
|---|---|---|---|---|---|---|
| Normal erythrocyte | 7.8 | 0.5 | 0.9447 | 85.1 | 2.19 | 8 |
| Microcyte | 5 | 0.4 | 0.9520 | 29.8 | 1.87 | 8 |
| Macrocyte | 10 | 0.65 | 0.8446 | 118.1 | 1.79 | 8 |
| Stomatocyte $(+,-)$ | 7.8 | (0.05, 0.95) | (0.9490, 0.4720) | 85.1 | 2.29 | 8 |

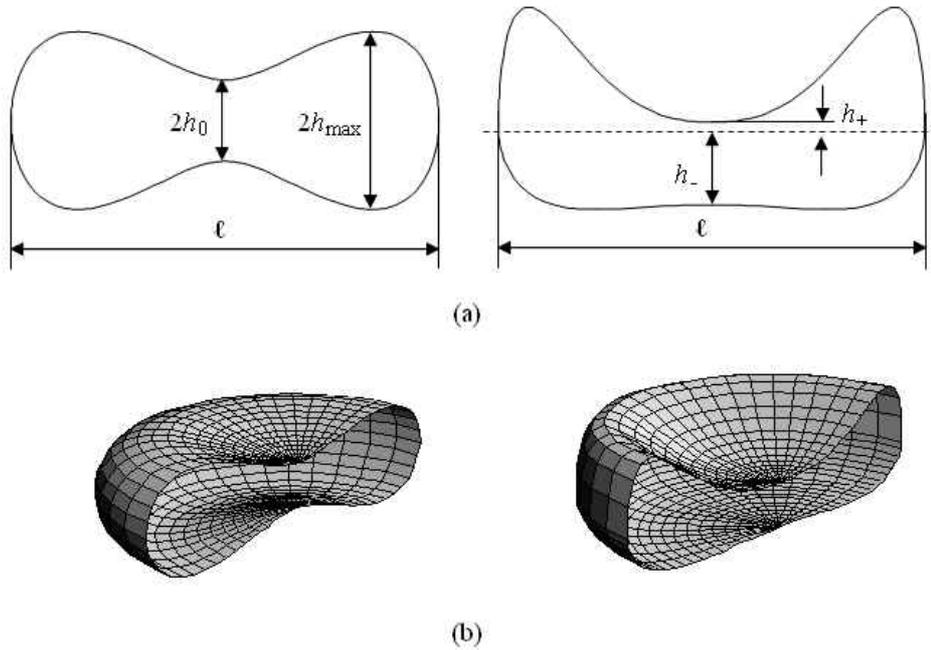

**Fig. 1 (Left)** Normal biconcave erythrocyte of diameter $\ell = 7.8\mu m$, maximum height $2h_{max} = 2.18\mu m$, **height at the center** $2h_0 = 1\mu m$ **and volume** $V = 85(\mu m)^3$ **which corresponds to a value of** $m = 0.9447$ **in Eq. 1. (Right) Stomatocyte of the same volume, height at the center, and diameter modeled with parameters** $\ell = 7.8\mu m$, $h_+ = 0.05\mu m$, $m_+ = 0.949$, $h_- = 0.95\mu m$ **and** $m_- = 0.472$ **in Eq. 2.**

## APPLICATION: TRANSMEMBRANE VOLTAGE IN A RF FIELD

To illustrate the practical use of these parameterizations we present a numerical calculation of the induced transmembrane potential in normal erythrocytes, macrocytes, microcytes and stomatocytes exposed to a 1800 MHz plane wave, a typical carrier frequency in cellular phones and in wireless surveillance systems. The transmembrane potential is a basic magnitude to understand the relation between the exposition to radiofrequency fields and the subsequent physiological and morphological reactions at cell level [Gedde and Huestis, 1997].

In our setup, the cells are immersed in an external continuous medium (the radiation region) formed by electrolytes in free water. The electric relative permittivity and conductivity of the external medium are those of the physiological saline, $\varepsilon_r = 80$ and $\sigma = 0.12 \text{Sm}^{-1}$ respectively. The cytoplasm is, in both cells, a physiological saline solution with a protein volume fraction of 0.26, relative permittivity $\varepsilon_r = 50$ and conductivity $\sigma = 0.53 \text{Sm}^{-1}$. The membrane is a shell of constant thickness $\delta = 8 \text{nm}$ with a frequency-independent relative permittivity $\varepsilon_r = 9.04$ and a very low conductivity $\sigma = 10^{-6} \text{Sm}^{-1}$. These values are typical for the erythrocyte structure and have been extensively used in the literature [Simeonova et al, 2002].

We already mentioned the advantages of an explicit parametric representation: in particular we can easily generate the uniform shell representing the membrane by shifting the parametric surface given by Eqs. 1 and 2 by a constant distance $\delta$ along the outer normal at each point, i.e.,

$$\mathbf{r}_\delta(u,\phi) = \mathbf{r}(u,\phi) + \delta \mathbf{n}(u,\phi) \tag{4}$$

where

$$\mathbf{n}(u,\phi) = \frac{(\partial \mathbf{r}/\partial \phi) \times (\partial \mathbf{r}/\partial u)}{\|(\partial \mathbf{r}/\partial \phi) \times (\partial \mathbf{r}/\partial u)\|}. \tag{5}$$

To determine the induced transmembrane voltage, we have applied a finite element numerical technique, whereby we solve the full Maxwell equations using a discretization of the cell model geometry into tetrahedral elements with an adaptive mesh. For normalizing purposes the electric field intensity is set to 1 V/m [Muñoz et al, 2004].

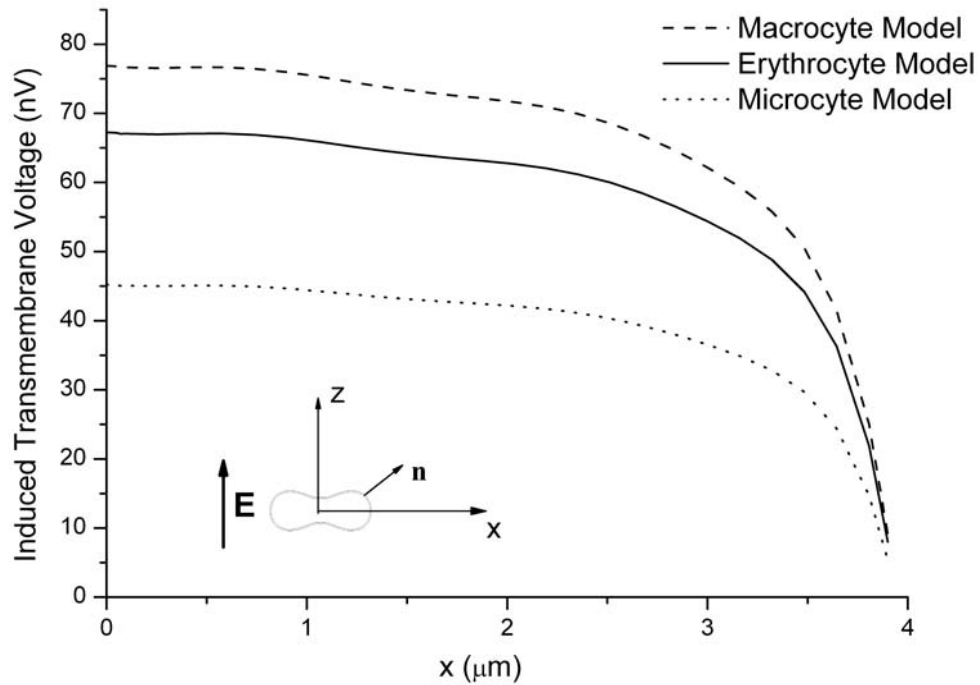

**Fig. 2** Induced transmembrane potential in a microcyte, a macrocyte and a standard erythrocyte as a function of the distance *x* along the major axes of the cells. The cells are exposed to an RF field of frequency 1800 MHz. The electric field has an intensity of 1 V/m and is linearly polarized parallel to the minor axes of the cells.

Figure 2 shows a comparison of the induced transmembrane potential in a microcyte and a macrocyte with the induced transmembrane potential in a standard erythrocyte as a function of the distance *x* along the major axes of the cells [Muñoz et al, 2005]. The polarization of the incident electric field is parallel to the minor axis of the cell. Note that the transmembrane potential is always significantly reduced as the electric field and the normal **n** to the membrane surface become perpendicular. This can be explained by the stronger mutual polarization in the neighboring regions of the membrane along the minor cell axis. Figure 2 also shows that the transmembrane voltage induced in the macrocyte and in the microcyte are always higher and lower respectively than the corresponding value for the normal erythrocyte cell. These results show the key role of the size of the cell on the induced transmembrane voltage.

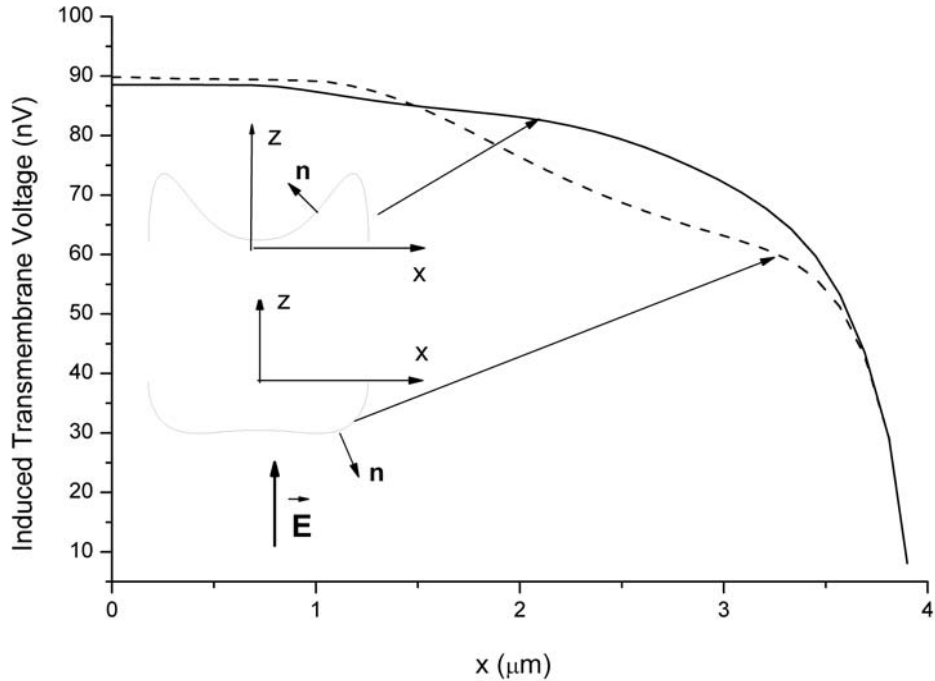

**Fig. 3 Induced transmembrane potential in the upper and lower parts of a stomatocyte as a function of the distance *x* along the major axis of the cell. The cells are exposed to an RF field of frequency 1800 MHz. The electric field has an intensity of 1 V/m and is linearly polarized parallel to the vertical axis.**

In contrast to the symmetrical erythrocyte shapes, the asymmetry of the stomatocyte around the major axis leads to different values of the induced transmembrane potential in the upper and in the lower parts of the cell. Fig. 3 shows this induced transmembrane potential both in the upper and in the lower parts of a stomatocyte as a function of the distance $x$ along the major axis. Again, the incident external electric field is parallel to the vertical axis. Note that in the flattest regions of the lower and upper parts of the stomatocyte the applied external field is parallel to the normal **n** to the membrane surface. As the values of the conductivities of both the external medium and the membrane are very low, the transmembrane voltage in these regions is, as expected, mainly governed by the ratio of the respective permittivities.

**CONCLUSIONS**

We finally point out that the common structure of the parameterizations given by Eqs. 1, 2a and 2b is

$$\mathbf{r}(u,\phi) = \left( \frac{\ell}{2} \operatorname{cn}(u,m) \cos\phi, \frac{\ell}{2} \operatorname{cn}(u,m) \sin\phi, h \operatorname{sn}(u,m) \left( \frac{\operatorname{dn}(u,m)}{\operatorname{dn}(U,m)} \right)^p \right) \qquad (6)$$

where the exponent $p$ (here equal to 1, 3 and 2 respectively) is a new parameter which in effect controls the gross features of the shape. In fact, taking into account that the Jacobi elliptic functions can be considered as deformations of the familiar trigonometric functions to which they reduce for $m = 0$ ($\operatorname{sn}(u,0) = \sin u$, $\operatorname{cn}(u,0) = \cos u$ and $\operatorname{dn}(u,0) = 1$, in which case the parametric equation (6) represents a revolution ellipsoid) and taking into account also the elementary properties of these Jacobi functions [Abramowitz and Stegun, 1972] it can be seen that the essential features of the parameterization (6) are maintained for positive (not necessarily integer) values of $p$. This new, continuous parameter can be used to advantage in the modeling not only of altered shapes, but of continuous deformations of the normal erythrocyte into these deformed shapes.